\documentclass[a4paper,12pt]{article}
\title{On bottom mixing with exotic quarks}
\author{ F. M. L. Almeida Jr and J. A. Martins Sim\~oes\\
Instituto de F\'{\i}sica\\
Universidade Federal do Rio de Janeiro\\
Ilha do Fund\~ao, Rio de Janeiro \\
BR-21945-970, RJ, Brazil}
\date{}
\begin{document}
\maketitle
\begin{abstract}
In this paper we present a calculation of the effects of bottom mixing with new exotic quarks in the forward-backward and left-right asymmetries, the bottom branching ratio and the QCD coupling constant. A global fit with the recent data on these quantities is done and stringent bounds are obtained. We discuss the effects of different isospin signatures for the new possible exotic quarks. The consequences for superstrig-inspired $E_6$ models are discussed. Constraints on the bottom mixing with the isosinglet quarks of the fundamental 27-plet are presented.
\end{abstract}
       Some extensions of the standard model predict the existence of new quarks and leptons. This is the case of the superstring-inspired $E_6$ models \cite{1}. In the fundamental $27$ representation we have new $Q = - 1/3$ isosinglet which can be mixed with the standard bottom quark \cite{2}. In $SO(n)$ we can also have mirror fermions \cite{3}. These models have gained a renewed interest with the recent possibility that neutrinos have non-zero mass \cite{4}. From general arguments, the present neutrino masses can be a hint to the physics at the grand unification scale at $10^{15}GeV$ \cite{5}. A natural scenario for the neutrino mass spectrum is the possibility of new isosinglet neutral leptons as indicated in the 27-plet. Other scenarios could also be possible and it is a fundamental problem in present day elementary particle physics to find evidences for the other consequences of any model which could generalize the standard model. However, these theoretical expectations have not received, so far, any direct experimental support \cite{6}. The validity of the standard model has been tested at the level of quantum corrections and this puts strong limits on new physics. New particles, if they exist, must be at a scale well above 100 GeV \cite{6,7}. It is well known that deviations from the standard model couplings must be necessary very small \cite{7,8}.\par
	 In this paper we suggest that an important step in the search for an extended model, such as $E_6$ is the possibility of new mixing in the quark sector. Of particular interest are the bounds involving the bottom quark. Since the third family has high mass states, there is a theoretical prejudice that deviations from the standard model will be more important here. In recent years, this seemed to be the case with the bottom asymmetries and hadronic branching ratio $R_b$. The more recent data has reduced these deviations to an acceptable level of less than $2\sigma$ \cite{6,7}.      In this paper we present a calculation of the effects of  bottom mixing with new exotic quarks in the  forward-backward and left-right asymmetries, the bottom hadronic branching ratio and  the QCD coupling constant. A global fit with the recent data on these quantities is done and stringent bounds are obtained.\par
       We consider three models, which differ by the $SU(2)\bigotimes U(1)$ assignment to new quarks and keep the same standard model attributes to the bottom:
\begin{enumerate}
\item{-} Vector singlet model (VSM). This is the case of the fundamental $27$ of $E_6$, with a new isosinglet quark, usually called  $"h"$.
\item{-} Vector doublet model(VDM). In this model, two new doublets are introduced, with left and right helicities.
\item{-} Fermion-mirror-fermion model (FMF). In this model we have a new doublet and singlet with opposite left and right assignments relative to the standard model.

\end{enumerate}
       In all models we allow different left and right mixing angles. It is then straightforward to calculate, at tree level the change in the  $Zbb$ couplings, given in table I, where $s_i^2=sin^2\theta_i$.\par
       We now discuss the general hypothesis involving our results. The first point to clarify is that we have taken into account in our calculation only the effects of mixing. In the class of models considered above one has many other effects such as new gauge bosons and scalars. A complete treatment of all the contributions is highly desirable but we will then face the problem of a large number of unknown parameters. It is known \cite{8} that there are no significant contributions of these new particles to radiative corrections and so we have decided to study only the mixing angle effects. This is equivalent to consider the weak isospin difference for each model.  We are also supposing that mixing effects are small, as shown by several authors \cite{9,10}. It is well known at present that radiative quantum corrections in the standard model are necessary in order to fit the experimental high precision data. With this in mind we consider that the changes in the physical observables due to new bottom mixing will make small contributions to the full standard model calculations, including first order corrections. So, we present the results of our calculation as powers of a small mixing angle and keep only the first term. They are displayed in equations 1 and 2. In equations 1, $"A"$ means the forward-backward and the left-right asymmetries. We call attention to the diference of signs in each expression. This is a direct consequence of the  isospin content of each extended model. Our calculation was performed in the on-shell scheme, with $sin^2\theta_W=0.2230$ and $m_t=174.3 $ GeV. In equations 2, for $R_b$, all models tend to reduce the standard model prediction. For the asymmetries, the VDM and VSM models show opposite corrections whereas the FMF model tends to cancel each correction.\par

\begin{eqnarray}
A_{VSM}&=&A_{SM}(1-0.1551 s_L^2)\nonumber\\
A_{VDM}&=&A_{SM}(1+0.8544 s_R^2)\\
A_{FMF}&=&A_{SM}(1+0.8544 s_R^2-0.1551 s_L^2)\nonumber
\end{eqnarray}

\begin{eqnarray}
R_{VSM}&=&R_{SM}(1-2.2768 s_L^2)\nonumber\\
R_{VDM}&=&R_{SM}(1-0.4132 s_R^2)\\
R_{FMF}&=&R_{SM}(1-0.4132 s_R^2-2.2768 s_L^2)\nonumber
\end{eqnarray}

       We have also performed a global fit to the present experimental data. The fit is compatible with zero mixing for all models.We have obtained, at $95\%$ confidence level,
the following upper bounds for each model 
\begin{eqnarray}
s_L^2<0.046 & \textrm{for VSM} \nonumber\\
s_R^2<0.002 & \textrm{for VDM} \\
s_R^2<0.087,\hspace{.5cm} s_L^2<0.046 & \textrm{for FMF}\nonumber
\end{eqnarray}
The numerical results for the data are shown in table II, for the Particle Data Group average in their 1999 update. They clearly show no evidence for mixing within the models discussed. We have checked the change in $\alpha_s$ due to the bottom mixing. There is a correlation between $\alpha_s$ and the changes in $Zbb$ [11]. With the upper bounds above there is a small contribution, bounded by $\Delta\alpha_s < 0.005$.\par
It is well known that $A^b$ is related to $A^{0,b}_{FB}$ and to $A_{leptonic}$\cite{6}.For the SLC measurements, we have $A_b=0.892\pm0.016$ and for the LEP results alone we have $A_b=0.904\pm0.018$. The global fit for these cases  shown no significate difference from the PDG average.\par

	In conclusion, the search for deviations from the standard model predictions in the bottom parameters can be a window to test the predictions of grand-unified models. As there is an enormous experimental activity on b-physics, it is expected that the uncertainties on the basic b-parameters will be strongly reduced very soon \cite{12}. If the future data is closer to the standard model predictions, our results will imply new stronger bounds on mixing angles. If, on the contrary, there are significant experimental discrepancies with the standard model, our calculation could be very useful in establishing the theoretical origin of then. For example, if $R_b$ turns to be above the standard model prediction, then this effect can not be attribute to mixing with new quarks in any of the models here considered. For the asymmetries, one must look in equation (1) for a correct sign in any possible deviation. In particular, for the fundamental $27$ in $E_6$, with a new isosinglet exotic quark, one expects that the experimental asymmetries should be smaller than the standard model predictions.\par
	
{\it Acknowledgments:} This work was partially supported by the following Brazilian agencies: CNPq, FUJB, FAPERJ and FINEP.

\begin{table}[1]
\begin{center}
\begin{tabular}{|c|c|c|c|}
\hline
& & & \\
 & VSM & VDM & FMF\\
& & & \\
\hline
& & & \\
$g_v$ & $g_v^{SM}+\frac{1}{2}s_L^2$ & $g_v^{SM}-\frac{1}{2}s_R^2$ & $g_v^{SM}-\frac{1}{2}s_R^2+\frac{1}{2}s_L^2$\\
& & & \\
\hline
& & & \\
 $g_a$ & $g_a^{SM}+\frac{1}{2}s_L^2$ & $g_a^{SM}+\frac{1}{2}s_R^2$ &  $g_a^{SM}+\frac{1}{2}s_R^2+\frac{1}{2}s_L^2$\\
& & & \\
\hline
\end{tabular}
\end{center}
\caption{New vector and axial-vector couplings for the $Zbb$ vertex}
\end{table}

\begin{table}[2]
\begin{center}
\begin{tabular}{|c|c|c|c|c|c|}
\hline
& & & & & \\
 & SM & Experimental & VSM & VDM & FMF\\
& & & & & \\
\hline
& & & & & \\
$A^b$ & $0.9348\pm 0.0001$ & $0.911\pm 0.025$ & $0.9345$ & $0.9348$ & $0.9404$\\
& & & & & \\
\hline
& & & & & \\
$A^{FB}$ & $0.1034\pm 0.0009$ & $0.0988\pm 0.0020$ & $0.1034$ & $0.1034$ & $0.1040$\\
& & & & & \\
\hline
& & & & & \\
$R^b$ & $0.2158\pm 0.0002$ & $0.2164\pm 0.0007$ & $0.2148$ & $0.2158$ & $0.2141$\\
& & & & & \\
\hline
\end{tabular}
\end{center}
\caption{Assymetries and hadronic branching ratio for the standard model(SM), experimental(Exp.) values and their 95\% confidence level limits for the different models. The experimental data is from the Particle Data Group update of November, 1999.}
\end{table}

\vfill
\end{document}